\documentstyle [preprint,aps,multicol, epsfig]{revtex}
\tightenlines

\begin{document}

\bibliographystyle{prsty}

\draft

\title{Current Path Properties of the Transport Anisotropy at Filling
Factor 9/2}

\author{R.L. Willett,  K.W. West, L.N. Pfeiffer}

\address{Bell Laboratories, Lucent Technologies, Murray Hill, NJ 07974}

\maketitle

\begin{abstract}
To establish the presence and orientation of the proposed striped phase in ultra-high 
mobility 2D electron systems at $\nu =$~9/2, current path transport properties
are determined by varying the separation and allignment of current and voltage contacts.  
Contacts alligned orthogonal to the proposed intrinsic striped phase produce voltages 
consistant with current spreading along the stripes; current driven along the 
proposed stripe direction results in voltages consistent with channeling along 
the stripes.  Direct comparison is made to current spreading/channeling properties 
of artificially induced 1D charge modulation systems, which indicates the 9/2
stripe direction.\vspace{0.5in}
\end{abstract}

Since the demonstration of anisotropic transport at 9/2 filling
factor\cite {Du:99,Lilly:99}, considerable effort has been made
to understand the underlying physical mechanism. The 
fundamental finding of these measurements is a peak in the longitudinal 
magnetoresistance at 9/2 for current driven along the [1 $\bar{1}$ 0] 
direction, but a minimum there for current driven and voltage
measured across this direction.  The total anisotropy can be 
dramatic, with peak to minimum ratios at 9/2 large at low temperatures.
Similar results are observed at higher filling factors 11/2, 13/2, and 
15/2. These results were found only in high mobility systems
$\mu $ $>$ 10$^{7}$cm$^{2}$/V-sec, and at low temperatures ( $<$ 100mK ), 
suggesting electron correlations as the origin of the effect. 
These experimental results are supplemented by findings at 9/2 that include 
non-linear I-V\cite{Lilly:99} and induction of a peak in resistance
for large in-plane magnetic field\cite{Lilly2:99,Pan:99}.

The theoretical picture of this phenomenon has focused on the
formation of charged stripes\cite{Koulakov:96}, and has as its
basis Hartree-Fock theory. It is proposed that 
in the middle of the high Landau levels, energetics favor separation of 
the otherwise homogeneous 2D system into stripes of different filling 
factors: at 9/2 the 2D gas separates into a unidirectional charge 
density wave comprised of stripes of alternating 
filling factors 4 and 5. These stripes are presumed to orient along 
the [1 1 0] direction\cite{Koulakov:96,MacDonald:00}, resulting in a 
resistance  peak for current driven and voltage measured across this 
direction, and a resistance minimum at 9/2 for current and 
voltage along this direction.  Away from 9/2 it has been further 
proposed that the stripes may break in a bubble phase, with one integral 
filling forming a bubble within a background of the next near integral 
filling factor. This Hartree-Fock picture has gained substantial support in 
numerical studies\cite{Rezayi:99} that extract distinct peaks in both 
the static density susceptibility and the density-density correlation 
in the ground state, strongly suggesting charge density wave ordering.

To date, no measurements have specifically
demonstrated charge stripe formation. Re-entrant Hall 
resistance has been shown to occur around 9/2 filling\cite {Du:99}, supporting 
the picture of two phases coexisting.  The most specific experimental
indication of the picture of stripe phase formation is a measurement
of the reorientation of anisotropy in a square quantum well sample\cite{other}.  In this work, 
a theoretical calculation based upon unidirectional charge density waves
predicts the values of in-plane B-field needed to induce and then
re-orient anisotropy in magnetoresistance of 
a multiple electronic sub-band system.  The ability of this model
to accurately describe such a complex experimental process, in addition to 
the previous numerical work\cite{Rezayi:99}, indirectly indicates
the stripe phase formation.  

At present there exists some question as to which direction the 
proposed striped phase is alligned.  At 9/2, certain models predict that the stripes of 
filling factor 4 and 5 are alligned along the [1 1 0] direction\cite{Koulakov:96,MacDonald:00}. 
This orientation implies that the high resistance state is due to 
transport across the lines of alternating integral filling factor,
and the low resistance state is for current along the lines.  Other 
theories\cite{harvard:00}, based upon Chern-Simon constructions, predict 
an orientation of the stripes consistent with this.  
Also, a natural band anisotropy exists in AlGaAs/GaAs heterostructures\cite{Kroemer:99}, 
and it is proposed that the orientation of the phase separation
is determined by this intrinsic feature.  Given these purely theoretical descriptions,
the orientation of these phases is an open experimental
concern.

In this letter, we present new transport results on exceedingly high mobility
2D systems demonstrating magnetoresistance which is consistent with 
current spreading and channeling as heuristically expected in a picture 
of stripe phase formation. In addition, these results at high Landau 
levels are shown to be similar to the current path properties measured
in magnetotransport in artificial 1D charge modulated systems: 
by comparision of these two experimental systems the direction of the 
stripe formation is inferred.  These results provide a direct experimental 
connection between the proposed striped phase ground state and the magnetotransport 
properties of a charge striped system. 

A series of four samples from two different wafers were used in this
study.  Mobility ranges from 23$\times$10$^{6}$cm$^{2}$/V-sec to 
28$\times$10$^{6}$cm$^{2}$/V-sec, in excess of sample mobilities 
used in previous measurements addressing 9/2 transport\cite {Lilly2:99,Pan:99}.
Contacts to the 2D electron system were either diffused indium or lithographically 
defined and diffused nickel/gold/germanium. All contacts are less than 
300 $\mu $m at their widest dimension, with lithographically defined contacts
less than 50 $\mu $m. Standard lock-in measurement 
at low frequencies was used, and the samples were cooled in a dilution
or He3 refrigerator.

Figures 1(a) and 1(b) demonstrate the transport anisotropy found in these samples
of high quality.  In 1(a) the inset describes current and voltage contact
configurations, with the sample square about 5mm on each side and the
contacts separated on the edges by more than 1 mm. For constant current driven
along the [1 $\bar{1}$ 0] direction in the two different voltage tap 
configurations a peak is observed at half-integral filling for the high 
Landau levels ($\nu $ $>$ 4): this is as has been observed in previous transport 
studies. In Figure 1(b) the transport anisotropy is recognized in the black 
trace, where the current is driven across [1 $\bar{1}$ 0] through contacts near
the center of the sample sides. As observed previously, for transport
across [1 $\bar{1}$ 0] a distinct minimum is seen at half-integral filling
of the high Landau levels.  Using the same voltage contacts but
now moving the current source and drain to the same edge of the sample
produces a peak at 9/2 that is qualitatively similar to that observed for 
transport in the orthogonal direction. Note that the anisotropy is therefore
not apparent if transport is examined only along the sample edges. 
This is as expected for this contact topology along the single edge of the 
sample; the measured resistance should consist of a 
mixture of the microscopic resistivities in the form of 
$\sqrt {\rho _{xx}  \rho _{yy} }$, where in this model small contacts are 
located on an infinite edge\cite {Simon:01}.  Tests of the 9/2 anisotropy  
in this study involve systematically examining the magnetoresistance,
or voltage distribution, from a constant drive current  
for different current/voltage separations of the same topology as the black
traces in Figure 1.

By varying the relative positions of the current and voltage taps, the 
current distribution can be coarsely deduced.  This is particularly elucidating 
in a simple anisotropic conductor.  In assuming an ohmic yet anisotropic conductor,
the voltage measured at any contact pair is proportional to the current
that reaches that contact set, which is dependent upon the current distribution
as established by both the current source/drain positions and the intrinsic
anisotropic resistivity of the conductor.  In the set of transport traces 
shown in Figure 2 voltage is measured between contacts alligned along either the
[1 $\bar{1}$ 0] direction (a), or along the [1 1 0] direction (b), with
current contacts used for each trace systematically further away from the voltage
leads as shown in the inset.  Again, the sample used here is about 5mm long
on each side.  The data taken in the two orthogonal directions show what can be 
interpreted as current spreading for transport in one direction but relative 
current channeling in the orthogonal direction.  Figure 2a shows little variation 
in the peak at 9/2 for measurement configurations where the current and voltage
leads are progressively separated by more than a factor of three in distance. 
An interpretation of these results is that current driven 
in the [1 $\bar{1}$ 0] direction spreads orthogonally (along [1 1 0]) 
specifically for filling factor 9/2 such that the voltage drop along 
[1 $\bar{1}$ 0] is relatively independent of the proximity of the current source.  
This implies as well that at 9/2 current driven along [1 1 0] should 
be channeled along that direction with a substantial reduction in 
measured voltage for increasing spatial separation between current and voltage contacts.
This is consistent with the results in Figure 2b.  The peaks adjacent to 9/2 demonstrate
large changes as the separation of current and voltage contacts increases.  However,
specifically at 9/2 the voltage detected appears to be small (or zero) for all 
transport configurations.

To more closely examine the properties of this transport orthogonal to [1 $\bar{1}$ 0]
a configuration with closely spaced contacts was employed.  The schematic of Figure 3
shows a sample with four contacts spaced 0.5mm center to center on opposite sides 
of a high mobility structure with width of 5mm and crystallographic orientation 
as shown.  Variation in voltage detected for different
current to voltage contact distances is dramatic.  For current driven orthogonal 
to [1 $\bar{1}$ 0] but with only a 0.5mm separation between I and V contacts a large
peak is observed at 9/2, similar to that seen in magnetoresistance in the orthogonal
direction.  However, as the separation of current and voltage leads is increased
the voltage measured decreases substantially: at 1.0 mm separation a remnant peak
is observed at 9/2.  At 1.5mm separation no significant voltage is measured at 9/2, as
observed in the data of Figure 2.  For this sample orthogonal transport 
(current driven along [1 $\bar{1}$ 0]) showed results similar to Figure 2a.
These findings demonstrate a striking dependence of the voltage measured
upon the proximity to current source/drain for transport along [1 1 0],
but show that the voltage detected in the [1 $\bar{1}$ 0] direction for current in
that direction is roughly independent of the proximity to the current contacts.  

These results are consistent with the 9/2 state inducing substantial current spreading
along [1 1 0] for current driven along [1 $\bar{1}$ 0], and also inducing the complementary 
case of current channeling along [1 1 0] for current driven along that direction.
This current pattern is as expected for a simple anisotropy in magnetoconduction at
9/2 where a relatively high resistance path is observed along [1 $\bar{1}$ 0] and a less 
resistive path occurs along [1 1 0].  If a charge separation into a striped phase
is at the root of these findings, then examining the current pattern of an 
artificially induced striped system should reveal similar current path properties.  
Magnetotransport for an artificially striped 2D system\cite{Smet:98,Willett:99} is shown 
in Figure 4. This sample is a high mobility 2D heterostructure with a 1D charge modulation 
induced by a shallow etch of the sample surface in the pattern of parallel lines with a 
1.2 $\mu $m period. For current driven along the artificial lines (Fig. 4a) a large change is observed in 
measured voltage as current and voltage contacts are separated.  This is true throughout
the magnetic field spectrum, for both the fractional and integral quantum Hall regimes 
as shown in the insets.  By comparison, for current driven orthogonal to the artificial 
lines, a substantially smaller change is observed in the measured voltage as the 
current and voltage contact separation is increased (Fig. 4b).  In progressing from 
the smallest to largest current/voltage contact separations the ratio of measured 
voltage at filling factors 3/4 and 9/2 is about 10 for current driven along the 
artificial lines; for current across the lines this ratio is about 2.  This contrast 
between transport along or across the artificial lines is amplified at lower 
temperatures.  The data of Figure 4 clearly demonstrate
that charge lines, artificially induced here, produce lateral current spreading if the 
current is driven orthogonal to the lines, and produce current channeling if current is 
driven along the lines.  This same current distribution pattern is the essential finding 
in the higher Landau levels as shown in Figures 1-3, and by direct comparison to the 
transport in the artificial system it is inferred that the intrinsic lines of the 9/2 striped
phase form parallel to the [1 1 0] direction.

While the direction of the intrinsic striped phase lines can be derived from this 
comparison to the fabricated 1D charge modulation, it remains unclear what sets the
preferred direction of the intrinsic stripes along [1 1 0].

To summarize, current paths have been measured using varied current and voltage lead 
spatial separations on high mobility 2D systems in the high Landau levels where transport 
anisotropies have been observed, and these results are compared to magnetotransport in an 
artificially striped 2DES.  Specifically at the high Landau level half-filling,
current spreading is observed when current is driven along the [1 $\bar{1}$ 0] direction, 
indicating that a barrier to current propagation exists along the [1 1 0] direction
as may be attributable to the intrinsic stripe phase.  When current is driven along the 
[1 1 0] direction, current channeling is observed, wherein only small voltages are measured 
away from the current contacts.  These current path properties are shown to be similar to an  
artificially induced charge striped system at high magnetic fields, allowing deduction
of the stripe phase direction for the intrinsic, high Landau level states. 

We acknowledge useful discussions with S. Simon, D. Natelson, and M. Manfra.


\pagebreak

FIGURE CAPTIONS\vspace{1in}

Figure 1.  Transport along the edges of a high mobility (23$\times$10$^{6}$cm$^{2}$/V-sec) 
2D heterostructure (sample A) with contact configurations as shown in the inset to a).  
The sample dimension is ~5mm on each side. Temperature is 65mK.\vspace{0.5in}

Figure 2.  Transport in sample B ($\mu$ =23$\times$10$^{6}$cm$^{2}$/V-sec) 
with systematically larger separation of voltage and current contacts for the 
two orthogonal directions as shown in the inset.  Note that voltage measurements
with current driven across [110] show almost no variation, in contrast to current
driven orthogonally.  The sample is again 5mm X 5mm.  Temperature is 65mK.\vspace{0.5in}

Figure 3.  Transport in higher mobility 2DES (28$\times$10$^{6}$cm$^{2}$/V-sec, sample C) with
current driven along [110] and progressively larger separation of current and voltage
contacts.  The separation of nearest contacts is 0.5mm center to center, and the sample
width is 5mm.  Temperature is 65mK.\vspace{0.5in} 

Figure 4.  Effect of changing separation of current and voltage contacts in a 2DES with 
a 1.2~$\mu$m period density modulation along the lines as shown in the inset.  Shown in the 
insets are respective transport at low B-fields in these modulated samples. Current 
channeling for current driven along the imposed lines is apparent by the precipitous 
change in measured voltage throughout the magnetic field spectrum in the traces shown in a.
Labeled voltage scales are consistent between a and b.\vspace{0.5in}
Temperature is 290mK.

\pagebreak
\begin{figure}
\epsfclipon
\epsfxsize=12cm
\epsfbox{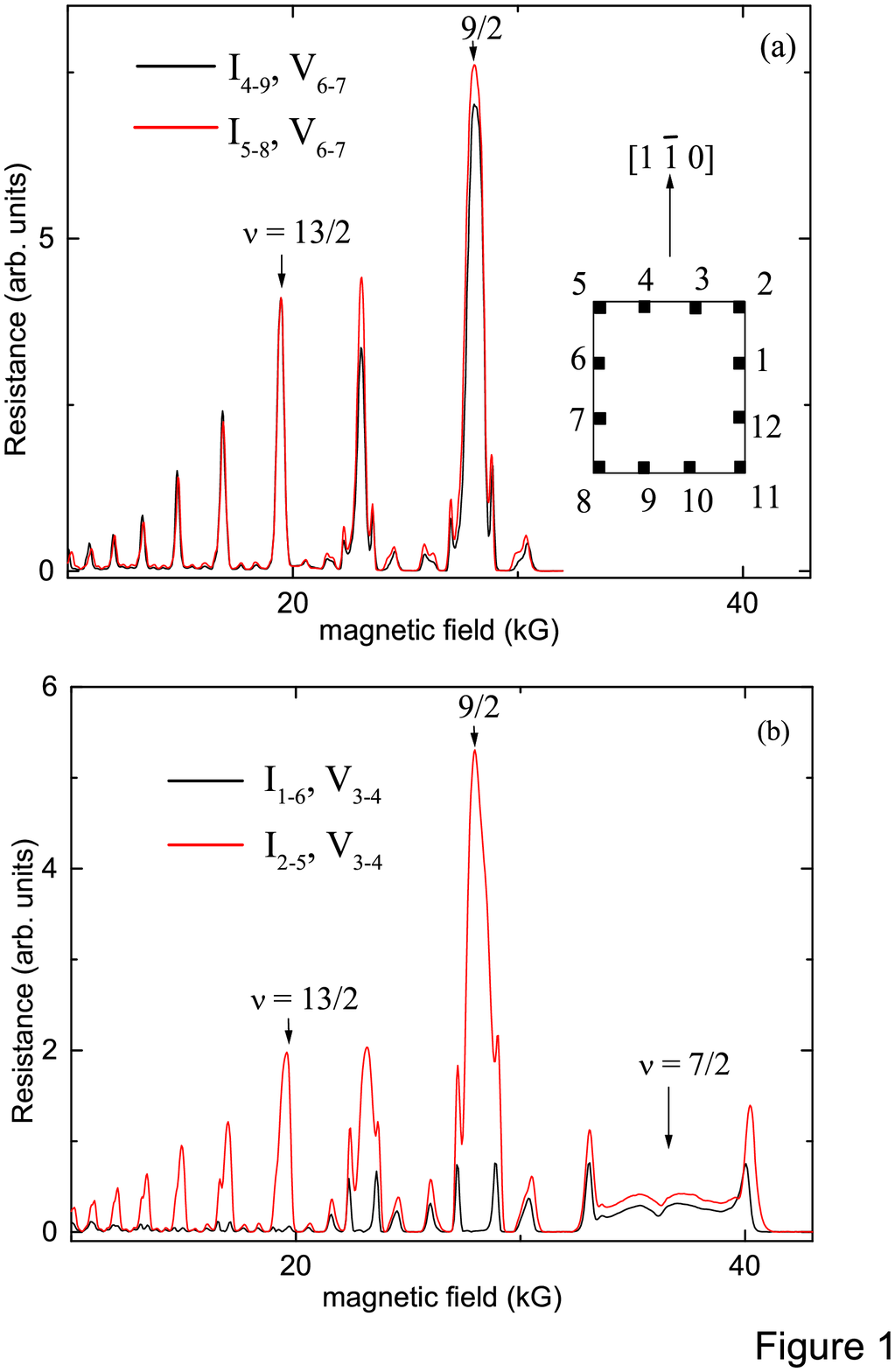}
\end{figure}

\pagebreak
\begin{figure}
\epsfclipon
\epsfxsize=12cm
\epsfbox{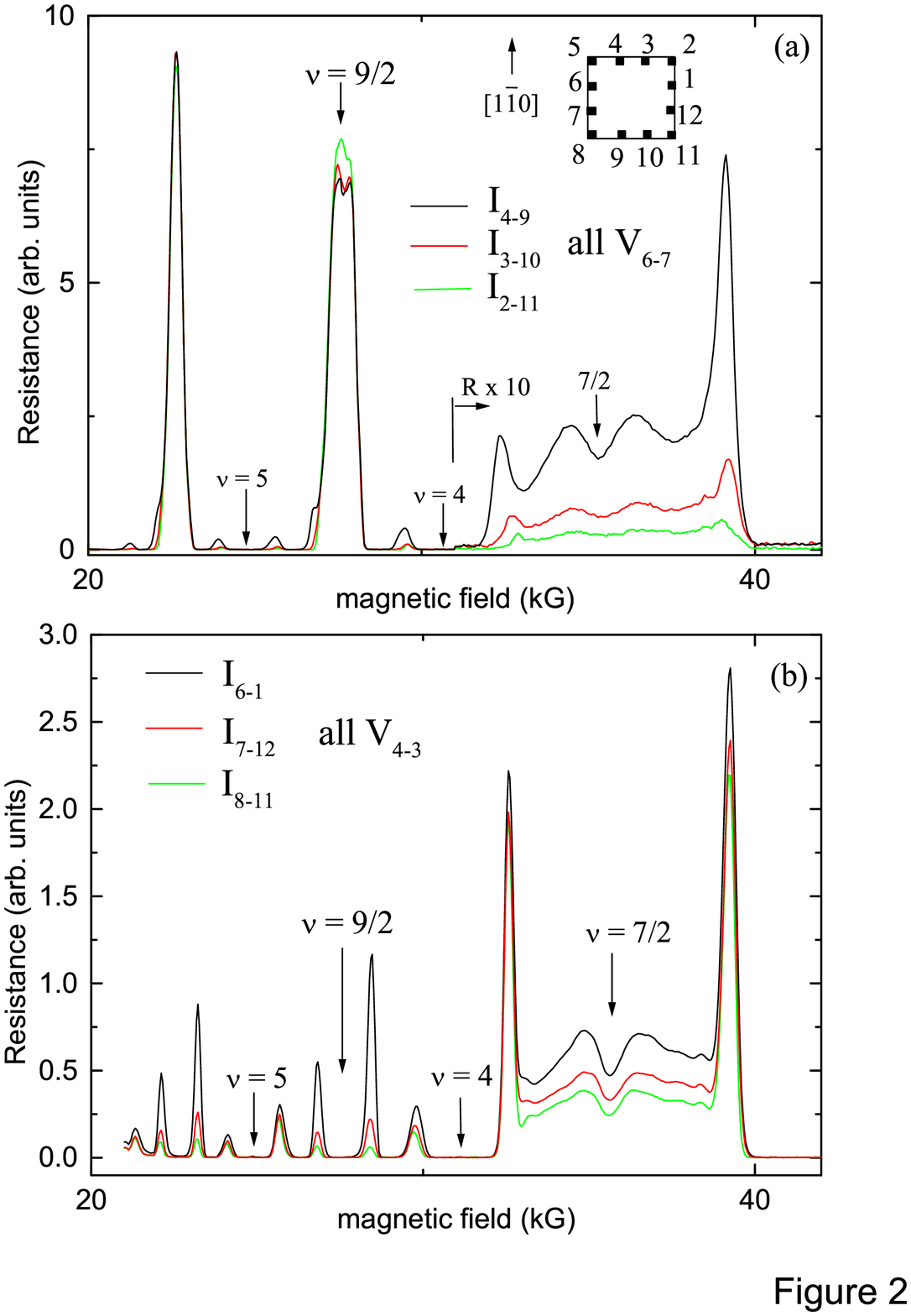}
\end{figure}

\pagebreak
\begin{figure}
\epsfclipon
\epsfxsize=12cm
\epsfbox{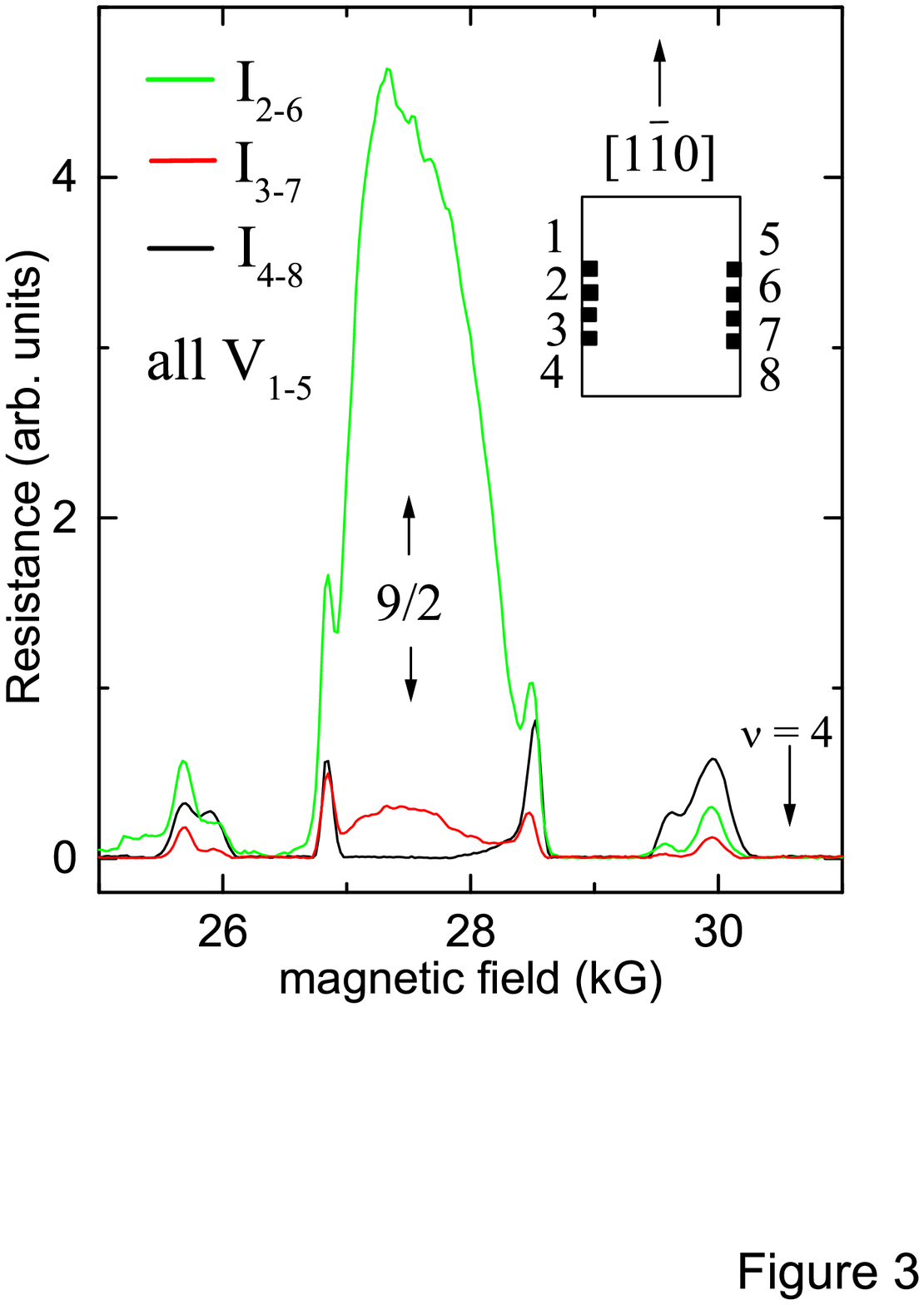}
\end{figure}

\pagebreak
\begin{figure}
\epsfclipon
\epsfxsize=12cm
\epsfbox{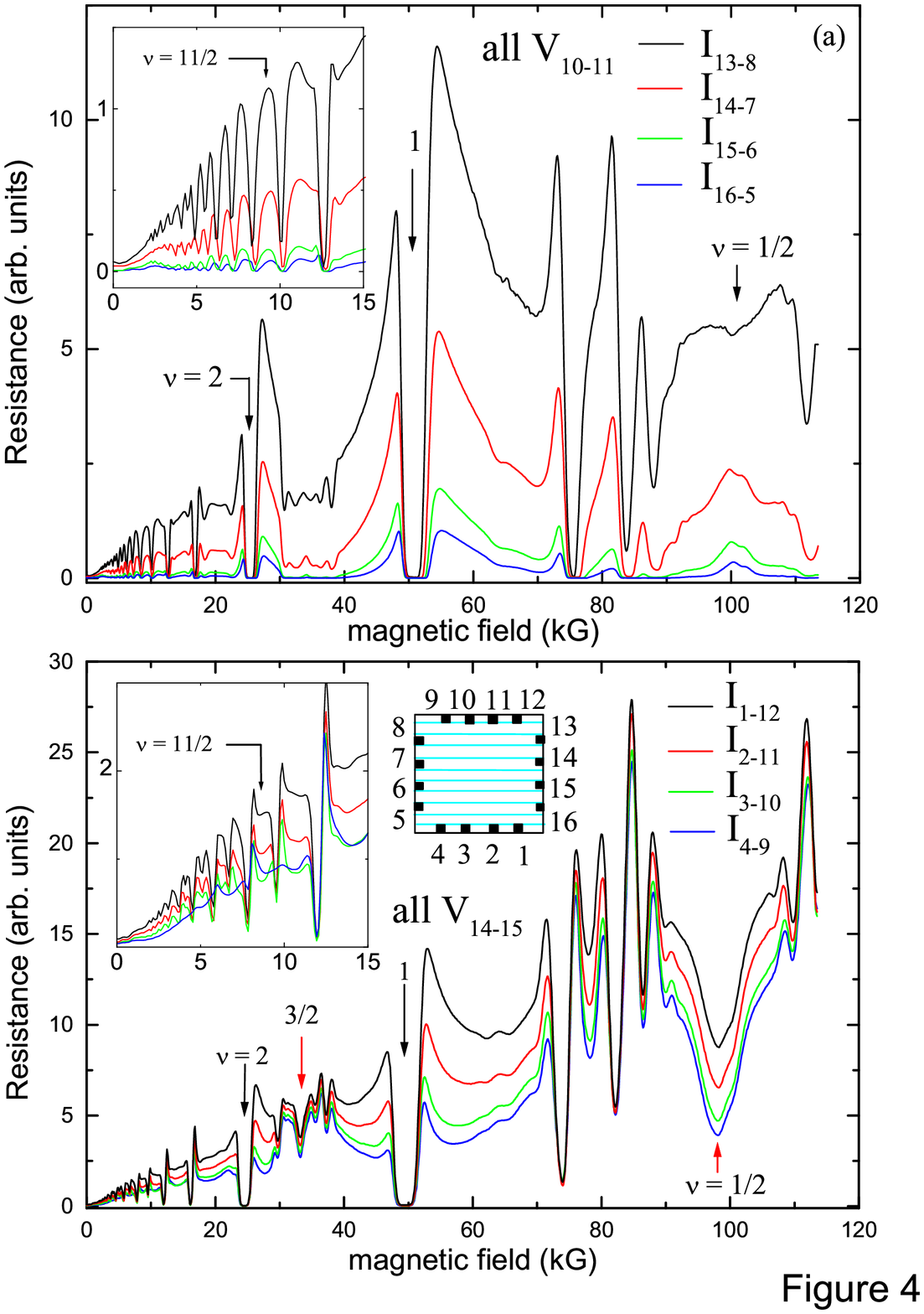}
\end{figure}

\end{document}